\documentclass{article}

\usepackage{amsmath,amssymb,amsfonts} 
\usepackage{graphicx,color} 
\usepackage{array} 
\usepackage[all]{xy}  

\usepackage{setspace} 

\newcommand{\np}{N_{+}} 
\newcommand{\nm}{N_{-}} 
\newcommand{\umin}{u_{min}} 
\newcommand{\umax}{u_{max}} 
\title{\bf{Systematic enumeration of configuration classes for entropic sampling
of Ising models}}

\author{BRUNO JEFERSON LOUREN\c{C}O~\footnote{brunojl@fisica.ufmg.br} \\ \smallskip
and \\ \smallskip
RONALD DICKMAN~\footnote{dickman@fisica.ufmg.br}
}

\begin{document}

   \maketitle
\begin{center}{\emph{
Departamento de F\'{\i}sica, Instituto de
Ci\^encias Exatas, and National Institute of Science and
Technology for Complex Systems,\\
Universidade Federal de Minas Gerais \\
C.P. 702, 30123-970, Belo Horizonte, MG, Brazil \\
}
}
\end{center}

\section*{\centering{Abstract}}
 
We describe a systematic method for complete
enumeration of configuration classes (CCs)
of the spin-1/2 Ising model in the energy-magnetization plane. 
This technique is applied to the antiferromagnetic Ising model
in an external magnetic field on the square lattice, which
is simulated using the tomographic entropic sampling algorithm.
We estimate the number of configurations, $\Omega(E,m,L)$,
and related microcanonical averages,
for all allowed energies $E$ and magnetizations $m$ for $L = 10$ to 30,
with $\Delta L = 2$.
With prior knowledge of the CCs, we
can be sure that all allowed classes are sampled 
during the simulation.
Complete enumeration of CCs also enables us to use the final estimate of $\Omega(E,m,L)$
to obtain good initial estimates, $\Omega_0(E,m,L')$, for successive system sizes ($L' > L$)
through a two-dimensional interpolation.
Using these results we calculate canonical averages of the thermodynamic quantities of interest
as continuous functions of temperature $T$ and external field $h$.
In addition, we determine the critical line in the $h$-$T$ plane using finite-size
scaling analysis, and compare these results with several approximate theoretical 
expressions.

{\bf{Keywords:}} Ising model; antiferromagnet; configuration classes; Monte Carlo simulation; phase diagram.



\section{Introduction}

The most important task of entropic sampling algorithms 
\cite{berg92a}$^{-}$\cite{dickman11a}
is to visit the full configuration space (CS) to obtain good estimates
of the number of configurations, $\Omega$, as functions of the energy E and
other quantities of interest.  
In studies of the Ising model in an external
field, for example, we require $\Omega(E,m)$ with $m$ the magnetization;
each allowed $(E,m)$ pair defines a {\it class} of configurations (CC).
Only if we know beforehand the possible values of $(E,m)$ for a given
system size, can we be sure that all CCs are sampled during the
simulation.

\begin{figure}[ph]
\centering
\includegraphics[width=0.8\textwidth]{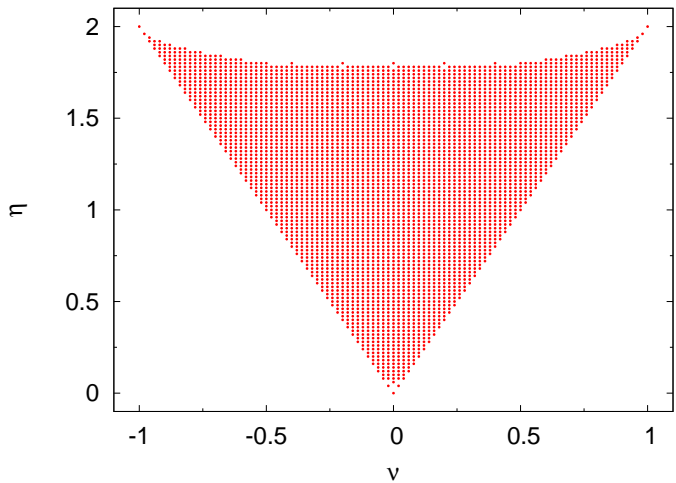}
\caption{\footnotesize{
Allowed configuration classes for system size $L = 10$; $\eta \equiv n / L^2$ and $\nu \equiv m / L^2$.}
}
\label{fig:d10b}
\end{figure}

Figure~\ref{fig:d10b} shows the CCs in the $n-m$ plane for the spin-1/2 Ising model
with nearest-neighbor (NN) interactions, on a square lattice of $L \times L$ sites
with periodic boundaries.
[We use $n$ to denote the number of NN
pairs of spins with the same orientation; the interaction energy
of the antiferromagnetic (AF) Ising model is $E = -2(L^2 - n)$.]   
Although the CCs tend to fill in a triangular region,
some gaps are evident near the lower apex and along the upper edge.
Knowing just which $(n,m)$ values are allowed for a given lattice size
is important if we are to implement entropic sampling with confidence.
In this paper we present a method for systematically enumerating all CCs
of Ising models in the $(n,m)$ plane.

We study the spin-1/2 antiferromagnetic Ising model in an external magnetic
field, whose energy is given by
\begin{equation}
 \mathcal{H} = - J \sum_{<i,j>} \sigma_i \sigma_j - h \sum_{i=1}^{N} \sigma_i = - E - h m,
\end{equation}
where $\sigma_i = \pm 1$, $<i,j>$ indicates a sum over NN
pairs of spins, $h$ is the external field, and $N$ is the number of spins;
the model is defined on a square lattice of $L \times L$ sites,
with periodic boundary conditions. 
Unlike the ferromagnetic Ising model ($J > 0$),
which exhibits a unique critical point in the $h-T$ plane and has an exact solution \cite{onsager44a},
the AF model ($J < 0$) possesses a critical line, which is not completely understood.

Various approximate methods have been applied to
determine the critical line of the AF Ising model on the square
lattice \cite{hartmann77a}$^{-}$\cite{penney03a};
these results, however, do not agree altogether.
Binder and Landau~\cite{binder80a} estimated the critical line via Monte Carlo
simulation, obtaining very good agreement with the approximate closed-form
expression of M\"uller-Hartmann and Zittartz~\cite{hartmann77a}, raising the possibility
that the latter expression was in fact exact. 
(It was later shown that this is not so \cite{wu90a}.) 
Hwang \emph{et al.} \cite{hwang07a} studied the AF Ising model
on the square and triangular lattices using a microcanonical transfer matrix
method and the Wang-Landau algorithm \cite{wang-landau01a}.
They performed an exact enumeration of the number of configurations, $\Omega(E,m)$,
and found CSs in the $E-m$ plane similar to that shown in Fig.~\ref{fig:d10b}.

Using the tomographic entropic sampling (TES) algorithm \cite{dickman11a} we estimate
$\Omega(n,m,L)$, and associated microcanonical averages, for lattice
sizes $L = 10$ to 30, with an increment $\Delta L = 2$. 
We then calculate the canonical averages of the thermodynamic quantities of interest.
Using these data we map out the critical line in the $h-T$ plane, and
compare our results with several theoretical expressions.

Prior determination of the set of allowed CCs is an important tool to
verify the quality of the sampling: we want to be sure that all CCs are
visited during the simulation. Since this algorithm uses an initial guess,
$\Omega_0(n,m,L)$, to begin the study, it is convenient to use the final
estimate $\Omega_N(n,m,L)$ (after the $N$-th iteration) to obtain the initial
guess $\Omega_0(n,m,L_0)$ of the next system size to be studied $(L_0 > L)$.
As we will show, good initial estimates, $\Omega_0(n,m,L)$, can be
obtained using a two-dimensional interpolation because we know a priori
the set of allowed CCs.

This paper is organized as follows. 
In Sec.~\ref{sec:cc} we define
the basic CCs and the respective allowed values of $(n,m)$
for the spin-1/2 Ising model on the square lattice;
the main goal of this section is to find 
all gaps in the $(n,m)$ plane.
In Sec.~\ref{sec:imp}, this information is used in
simulations of the AF Ising model via the TES algorithm. 
There we describe the method used to determine $\Omega_0(n,m,L_2)$ via two-dimensional
interpolation of the final estimate, $\Omega_N(n,m,L_1)$, of the previous
system size studied. 
Simulation results are reported in Sec.~\ref{sec:res}, for
the order parameter and the staggered susceptibility as functions of $h$
and $T$.
Points along the critical line in the $h-T$ plane are obtained using
finite size scaling analysis, and the results compared with several theoretical expressions.
We summarize our findings in Sec.~\ref{sec:conc}.

%
\section{Configuration Classes}
\label{sec:cc}

As pointed out above, one of the main problems in entropic sampling
methods is the prior determination of the complete set of configuration
classes for a given system size.
Let us denote by $N_+$ and $N_-$ the number of up and down spins,
respectively, on a square lattice  with $N = N_+ + N_- = L^2$ spins.
The number of pairs of opposite spins, $u$, and $N_+$ are
related to $n$ and $m$, respectively, via
\begin{equation}
 n = 2L^2 - u
\end{equation}
and
\begin{equation}
 m = 2 N_+ - L^2.
\end{equation}
Thus, once the possible values of $(\np, u)$ are determined, so are those of
$(n,m)$.  
Note that $u$, $n$ and $m$ can only take {\it even} values.
The gaps in CS fall near the maximum and minimum values
of $n$ ($n_{max}$ and $n_{min}$, respectively) for a given $m$. 
Therefore, we will
identify the possible values of $u$ near its maximum, $\umax$, and minimum,
$\umin$, for a given $\np \in [0,\frac{L^2}{2}]$; 
note that the number of configurations is
symmetric under interchange of $\np$ and $\nm$.

\subsection{Determining $u_{min}$ and nearby classes}

\subsubsection{Compact configurations}

\begin{figure}[ph]
\centering
\includegraphics[width=0.6\textwidth]{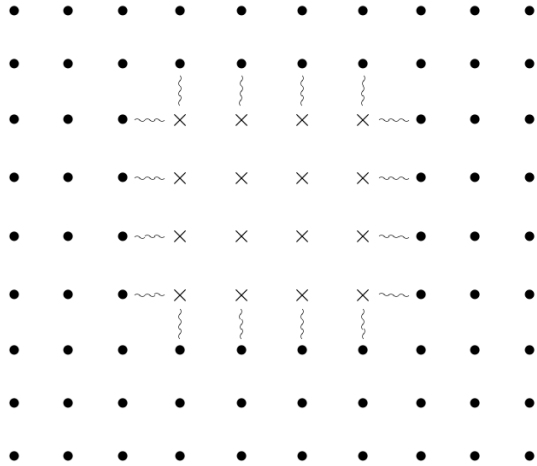}
\caption{\footnotesize{
Basic compact configuration. 
Up and down spins are represented by ``$\times$'' and 
``$\bullet$'', respectively; wavy lines
represent pairs of opposite spins.
}
}
\label{fig:comp1}
\end{figure}

\begin{figure}[ph]
\centering
\includegraphics[width=0.6\textwidth]{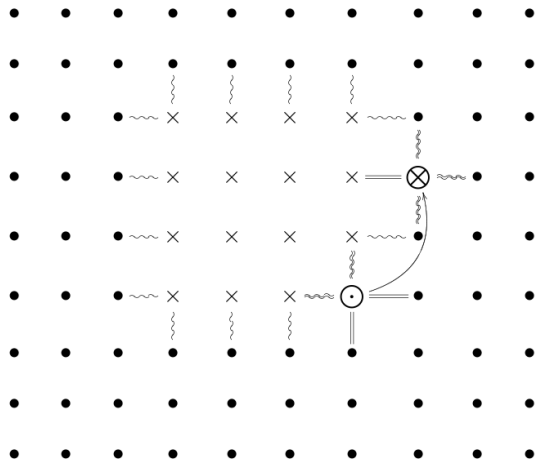}
\caption{\footnotesize{
Modified compact configuration. The new up and down spins are
represented by ``$\otimes$'' (previously ``$\bullet$'') and
``$\odot$'' (previously ``$\times$''), respectively;
double straight and double wavy lines represent new pairs of identical
and opposite NNs, respectively.
}
}
\label{fig:comp2}
\end{figure}

Compact configurations consist of a square or rectangular cluster
with $\np$ up spins. 
To begin, consider the case of a square cluster of size $l \times l$ $(2 < l < L)$ with $\np = l^2$,
as is illustrated in Fig.~\ref{fig:comp1}.
It is evident that this configuration corresponds to the
minimum value of $u$, $u^{(0)} = 4l$.
A configuration with the same
number of up spins, and $u = \np + 2$ is obtained by transferring an
up spin from one of the corners of the square to an edge, as shown in
Fig.~\ref{fig:comp2}.  
From this configuration, further rearrangements leading to
$u = \umin + 4$, etc., are possible.  
When $\np$ is not a square number, the
most compact configuration (i.e., with the smallest perimeter) is
a rectangle, or a square or rectangle with an incomplete layer of sites
along one edge.  
For $\np = l(l-1)$ we have $\umin = 4l-2$, while for
$l(l-1) < \np < l^2$, $\umin = 4l$.  
In all cases,
moving a corner site to an edge, one constructs a configuration with
$u = \umin + 2$, and further arrangements yield additional increases in $u$.

%
\subsubsection{Extended configurations}

Thus there are no gaps in the large-$n$ region due to compact configurations.
Such compact configurations, however, are not necessarily the ones that
minimize $u$ for a given value of $\np$.  
Consider for example the case $\np = k L$ with $k$ an integer.
The cluster of up spins can be arranged to wrap
around the lattice, yielding $\umin = 2 L$, independent of $k$.
We call such a configuration {\it extended}. 
For $k$ greater than a certain value,
on the order of $L/4$, the configurations that minimize $u$ are extended
rather than compact.
Suppose now that for a given $\np$, $\umin$ is obtained
with an extended configuration, and that all compact configurations
have $u$ strictly greater than $\umin$.
If $\np$ lies between $k L$ and $(k+1)L$,
then the configuration that minimizes $u$ has at least two corners,
and by the previous construction, configurations with $u = \umin + 2$ exist.
But if $\np = k L$, the minimizing configuration has no corners and any modification
yields a configuration with $u \geq \umin + 4$. 
This is how the gaps near the maximum values of $n$ arise.

Summarizing, if $L$ and $\np$ are such that
$\umin$ is obtained with a compact configuration,
then there are configurations with $\umin + 2$, $\umin + 4$, \ldots, etc., and no gap
exists.
If, on the other hand, $\umin$ is realized only for an extended
configuration, and $\np$ is an integer multiple of $L$, then there are
no configurations with $u = \umin+2$.

\subsection{Determining $\umax$ and nearby classes}

The largest possible value of $u$, $\umax$, occurs in a configuration such 
that $\np = L^2/2$ with spins arranged in a chess board (CB) configuration, such that
all up spins have down spins as NNs and vice versa.
One readily verifies that for $0 \leq N+ \leq L^2/2$, the maximum number of
unlike NN pairs is $\umax = 4 \np$.
Starting from the CB configuration, we can reduce $u$ by exchanging an up and a
down spin.  
If the exchanged spins are NNs, the resulting configuration
has $u = \umax - 6$; otherwise one has $u = \umax - 8$. 
Thus, for $\np = L^2/2$, there are no configurations with $u = \umax - 2$ or
$\umax - 4$.  
One readily verifies that configurations with $u = \umax - 10$, $\umax - 12$, etc.,
can be obtained via further exchanges of spins.

For $\np = L^2/2 -1$, we have $\umax = 2L^2 - 4$.
Such configurations can be
constructed by flipping one up spin in the CB configuration.
Starting from this configuration, exchanging a pair of spins, one can reduce $u$
by 4, 6 or 8, but there is no rearrangement which reduces $u$ by just two.
Thus for $\np = L^2/2 - 1$, there is no configuration having $u = \umax - 2$;
configurations with $u = \umax-4$, $\umax-6$, etc. do exist.

Finally, we note that for $1 < N+ < L^2/2 - 1$, there are no gaps in the
neighborhood of $u_{max}$.
To verify this, consider a configuration obtained by flipping $k$ up spins
in the CB configuration, so that
$\np = L^2/2 - k$,  where $1 < k < L^2/2$.
By hypothesis there are now at least four more sites with down spins
than with up spins. 
Starting from a configuration with $u = \umax$, in which
each up spin is surrounded by down spins, we can create a single NN pair
of up spins, with all six neighbors down, and with all remaining
up spins completely surrounded by down spins. 
In this manner, $u$ is reduced by two. 
Configurations with $u = \umin - 4$, $\umin - 6$, etc.
can be obtained by further exchanges of up and down spins.

Using the facts summarized above, it is straightforward to construct
an algorithm that determines which values of $u$ are possible, 
for a given $L$ and $\np$, 
and thereby which values $(n,m)$ are accessible for a given
system size. 
In the simulations reported below, we have verified that
our entropic sampling scheme converges to visit all allowed classes.

\section{Implementation}
\label{sec:imp}

Using tomographic sampling, we study the antiferromagnetic
Ising model in an external field on the square lattice;
we consider periodic boundary conditions and NN interactions.
The CCs of the systems are defined in the energy-magnetization
space $(n,m)$.

The TES method is applied in order to generate
estimates of $\Omega(n,m,L)$.
For the smallest system size $(L=10)$
we begin with a guess, $\Omega_0(n,m)$, obtained using a mean-field approximation.
For subsequent system sizes, however, we use a two-dimensional interpolation of
$\Omega_N(n,m,L)$ (the final estimate of $(n,m)$ for the smaller system size, $L$)
to construct $\Omega_0(n,m,L')$, where $L' > L$.
For most studies we use five iterations, each one with $N_{sim} = 10$ initial configurations, which
are simulated for $N_U = 10^7$ lattice updates or Monte Carlo steps.

Let us denote by $\Gamma (n,m)$ and $\Gamma' (n_0 = n + \Delta n, m_0 = m + \Delta m)$ 
the CCs that contain configurations $\mathcal{C}$ and $\mathcal{C'}$, respectively.
The simulation uses a single spin-flip dynamics, so that the possible
variations of $n$ and $m$ are $\Delta n = 0, \pm 2, \pm 4$ and  $\Delta m = \pm 2$.
At iteration $j$, the acceptance probability for the transition $(\mathcal{C} \to \mathcal{C'})$ is
\begin{equation}
\label{eq:p_trans}
p_j(\Gamma \to \Gamma') = \textrm{min} \bigg[ \frac{\Omega_{j-1}(\Gamma)}{\Omega_{j-1}(\Gamma')}, 1 \bigg].
\end{equation}
These probabilities are stored in a table. For each configuration generated,
be it a new one (if it is accepted) or the same (if it is rejected),
we update the sums used to calculate the microcanonical and
canonical averages of $\phi$, $|\phi|$, $\phi^2$, and $\phi^4$, where
\begin{equation}
\label{eq:phi}
\phi \equiv m_A - m_B
\end{equation}
is the order parameter (staggered magnetization); 
$m_{A,B}$ are the magnetizations of the two sublattices.
At the end of each iteration $j$
the estimate of $\Omega_j(n,m)$ is refined according to
\begin{equation}
\label{eq:at_ohm}
\Omega_{j}(\Gamma) = \frac{H_{j}(\Gamma)}{\overline{H}_{j}(\Gamma)} \, \Omega_{j-1}(\Gamma),
\end{equation}
where $H_j(\Gamma)$ is the histogram containing the number of times
the CC $\Gamma$ is visited during the sampling, and $\overline{H}_j(\Gamma)$
is the average of $H_j(\Gamma)$ over all accessible CCs;
the acceptance probability 
[Eq.~(\ref{eq:p_trans})] is updated using the new estimate of
$\Omega_j(n,m)$ and the histogram, $H_j(n,m)$, is set to zero.

A single iteration of our method consists of ten independent simulations,
each involving $10^7$ lattice updates, and each beginning from a different
initial configuration.  (By a lattice update we mean one attempted flip
per spin; the initial configurations include, high, low, and intermediate
interaction energies and both signs of the magnetization.)

\subsection{Determining $\Omega_0(n,m,L)$: mean field approach}

As noted above, for the smallest system size we use an estimate of
$\Omega_0(n,m,L)$ obtained via a mean-field approximation, specifically
\begin{equation}
\label{eq:mean_f}
\frac{1}{L^2} \ln \Omega_0(n,m,L) = \frac{1}{L^2} \ln \Omega(m) - \frac{(n - \langle n \rangle)^2}{2L^2 \sigma^2} - \frac{\ln \sigma}{L^2} + const. \, ,
\end{equation}
where $\sigma \equiv \sigma(m) = \sqrt{\textrm{var}(n)}$, and
\begin{equation}
\frac{\ln \Omega(m)}{L^2} \simeq \ln 2 - \frac{1}{2}[(1 + m/L^2) \ln (1 + m/L^2) + (1 - m/L^2) \ln (1 - m/L^2)].
\end{equation}

To obtain this expression, we first note that
$\Omega(m, L) = \sum_n \Omega(n,m,L) =  \binom {L}{\np}$,
and use Stirling's approximation. 
The dependence on $n$ is then obtained by estimating $\langle n \rangle$
and $\textrm{var}(n)$ for a given $m$ and $L$, using a random-mixing approximation, and
supposing that $n$ follows a Gaussian distribution.

\begin{figure}[ph]
\centering
\includegraphics[width=0.8\textwidth]{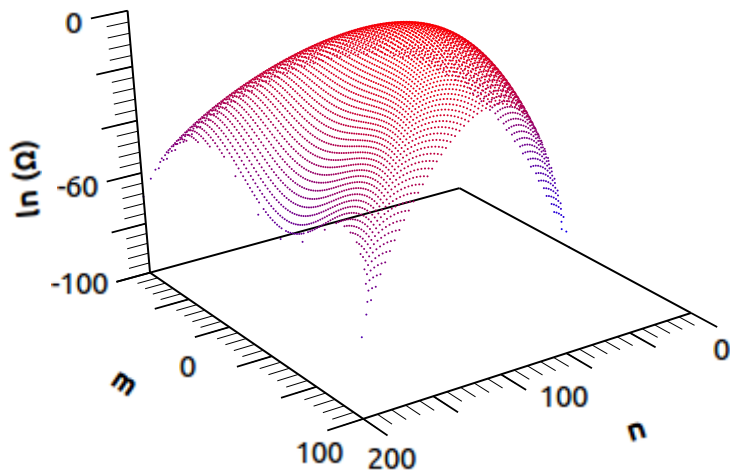}
\caption{\footnotesize{
Estimate of $\ln \Omega_0(n,m,L=10)$ via mean-field approximation.}
}
\label{fig:ohm2}
\end{figure}

\begin{figure}[ph]
\centering
\includegraphics[width=0.8\textwidth]{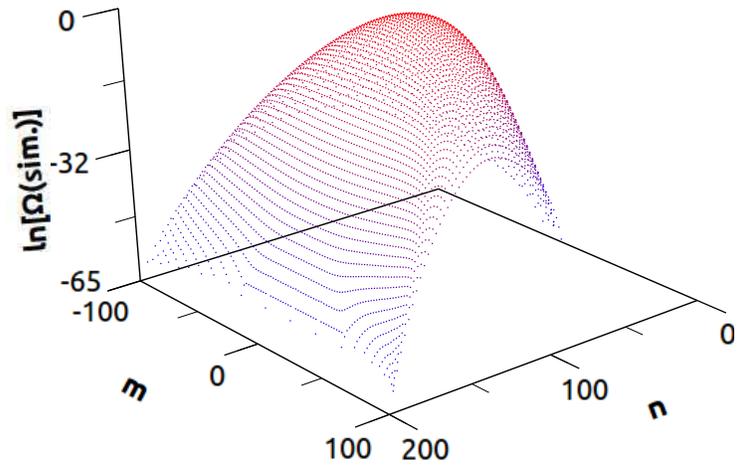}
\caption{\footnotesize{
Final estimate of $\ln \Omega_5(n,m,L=10)$ after the 5$th$ iteration of the simulation.}
}
\label{fig:ohm10s}
\end{figure}

\begin{figure}[ph]
\centering
\includegraphics[width=0.8\textwidth]{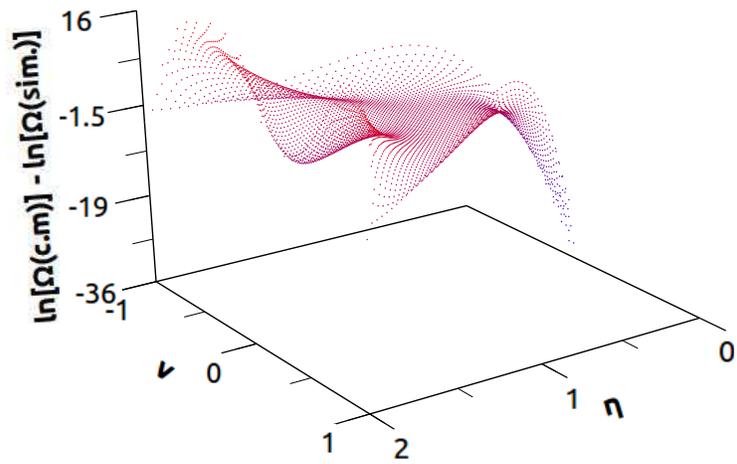}
\caption{\footnotesize{
Difference between \mbox{$\ln \Omega_0(n,m,L=10)$}, 
estimated
via mean-field approximation, and the final
simulation result after the 5th iteration, $\ln \Omega_5(n,m,L=10)$.}
}
\label{fig:dif_cm5}
\end{figure}

In Fig.~\ref{fig:ohm2} we plot the estimate of $\ln \Omega_0(n,m,L=10)$ given by Eq.~(\ref{eq:mean_f}).
We present in Fig.~\ref{fig:ohm10s} the final estimated value of $\ln \Omega_5(n,m,L=10)$ after
the 5$th$ iteration of the simulation;
this result is quite similar to that presented by Hwang \emph{et al.}~\cite{hwang07a}
for the square lattice, which was obtained using Wang-Landau sampling \cite{wang-landau01a}.
We note that the differences between the initial estimate obtained via mean field
approximation and the final simulation result of $\ln \Omega(n,m,L=10)$ 
are more evident near the edges of the CS, specially near to the maximum values of $n$.
To have a better idea of how close this initial estimate is to the final simulation
result, we plot in Fig.~\ref{fig:dif_cm5} the difference between that estimate and the
final result of the simulation for $L = 10$.
Analyzing this figure, it is again clear that the differences are larger near the edges of the CS.

%

\subsection{Determining $\Omega_0(n,m,L)$: interpolation}

\begin{figure}[ph]
\centering
\includegraphics[width=0.8\textwidth]{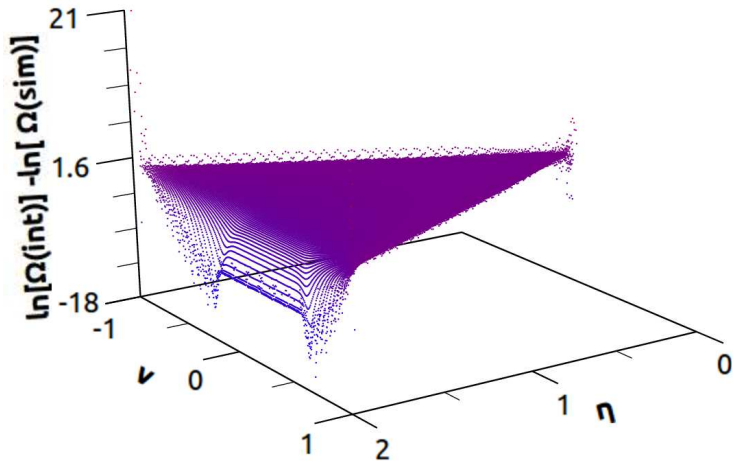}
\caption{\footnotesize{Difference between \mbox{$\ln \Omega_0(n,m,L=18)$}, 
estimated
by interpolating the final result of $L = 16$, and the final
simulation result after the 5th iteration, $\ln \Omega_5(n,m,L=18)$.}
}
\label{fig:dif_int5}
\end{figure}

We expect that the closer the estimate $\Omega_0(n,m)$ is to the (unknown) exact
value, $\Omega_e(n,m)$, the faster the simulation will converge, and the more
accurate our final estimate will be. 
Since the mean-field estimate worsens as the system size grows,
we only use it for the smallest system size
studied; initial estimates of subsequent system sizes are obtained by
interpolating the final estimate, $\Omega_N(n,m,L)$, of the previous system
size studied. 
Our procedure is based on the existence of the limiting
microcanonical entropy density as a function of the intensive parameters
$\eta$ and $\nu$ (the bond and magnetization densities, respectively),
\begin{equation}
s(\eta,\nu) = \lim_{L \to \infty} \frac{1}{L^2} \ln \Omega (n_L,m_L,L),
\end{equation}
where $n_L \simeq \eta L^2$ and $m_L \simeq \nu L^2$.  (Since $n$ and
$m$ are restricted to even integers we have $n_L = \eta L^2 + {\cal O}(1/L^2)$
and similarly for $m_L$.)  The idea is then to write
\begin{equation}
\label{eq:int2}
\frac{1}{L'^2} \ln \Omega_0 (n', m', L')
     = \frac{1}{L^2} \ln \Omega_N (\eta L^2, \nu L^2, L),
\end{equation}
where $\eta = n'/L'^2$, $\nu = m'/L'^2$, and the r.h.s. is evaluated by
extending $\ln \Omega_N$ to noninteger $n$ and $m$ via extrapolation and
interpolation.
Using this approach, we obtain better estimates, as is shown in
Fig.~\ref{fig:dif_int5}.  (Note that the largest differences continue to fall along the
edges of the CS.)

%

\subsection{Extrapolation and interpolation}

\begin{figure}[ph]
\centering
\includegraphics[width=0.7\textwidth]{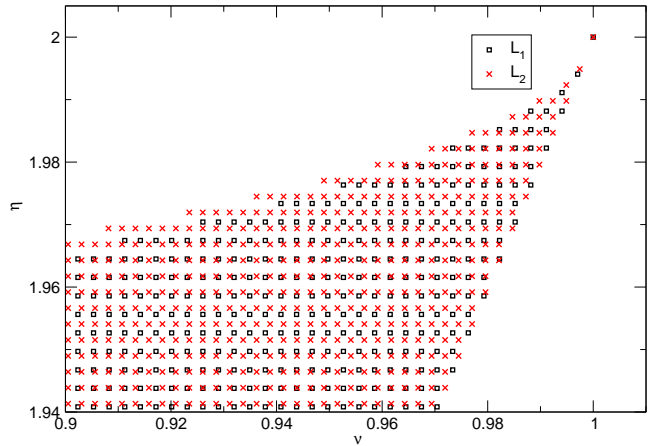}
\caption{\footnotesize{
Region of the  $(\eta,\nu)$ plane with every possible
configuration classes for system sizes $L_1 = 26$ and $L_2 = 28$.}
}
\label{fig:dohm1}
\end{figure}

Suppose we wish to construct the initial estimate $\Omega_0(n,m,L_2)$ on the
basis of the simulation results for a smaller system, $\Omega_N(n,m,L_1)$.
We could do this via {\it interpolation} if every CC
of the larger system were surrounded by four points of the smaller
one in the $\eta$-$\mu$ plane. 
Fig.~\ref{fig:dohm1} shows, however, that
along the edges of the CS, the points corresponding
to classes of the larger system are not surrounded by four points of
the smaller one.
For those points, one could in principle use
extrapolation rather than interpolation.  We found, however, that
direct extrapolation yields poor estimates for $\Omega$.

\vspace{1cm}

\begin{figure}[ph]
\centering
\includegraphics[width=0.7\textwidth]{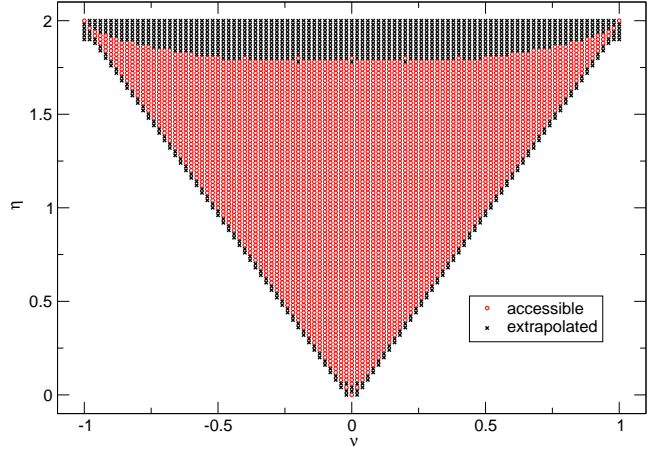}
\caption{\footnotesize{
Accessible and extrapolated configuration classes
of a system of size $L = 10$.}
}
\label{fig:de10}
\end{figure}

We obtain better estimates by first extrapolating the points along the
edges of the CS for the smaller system, such that every
point of the larger system
is surrounded by four points of the smaller.
Figure~\ref{fig:de10} shows the CS
with accessible and extrapolated CCs for $L = 10$.
Following this extrapolation we
perform a linear two-dimensional interpolation as per Eq.~(\ref{eq:int2}).

%

%
\section{Results}
\label{sec:res}

\begin{figure}[!h]
\centering
\includegraphics[width=0.8\textwidth]{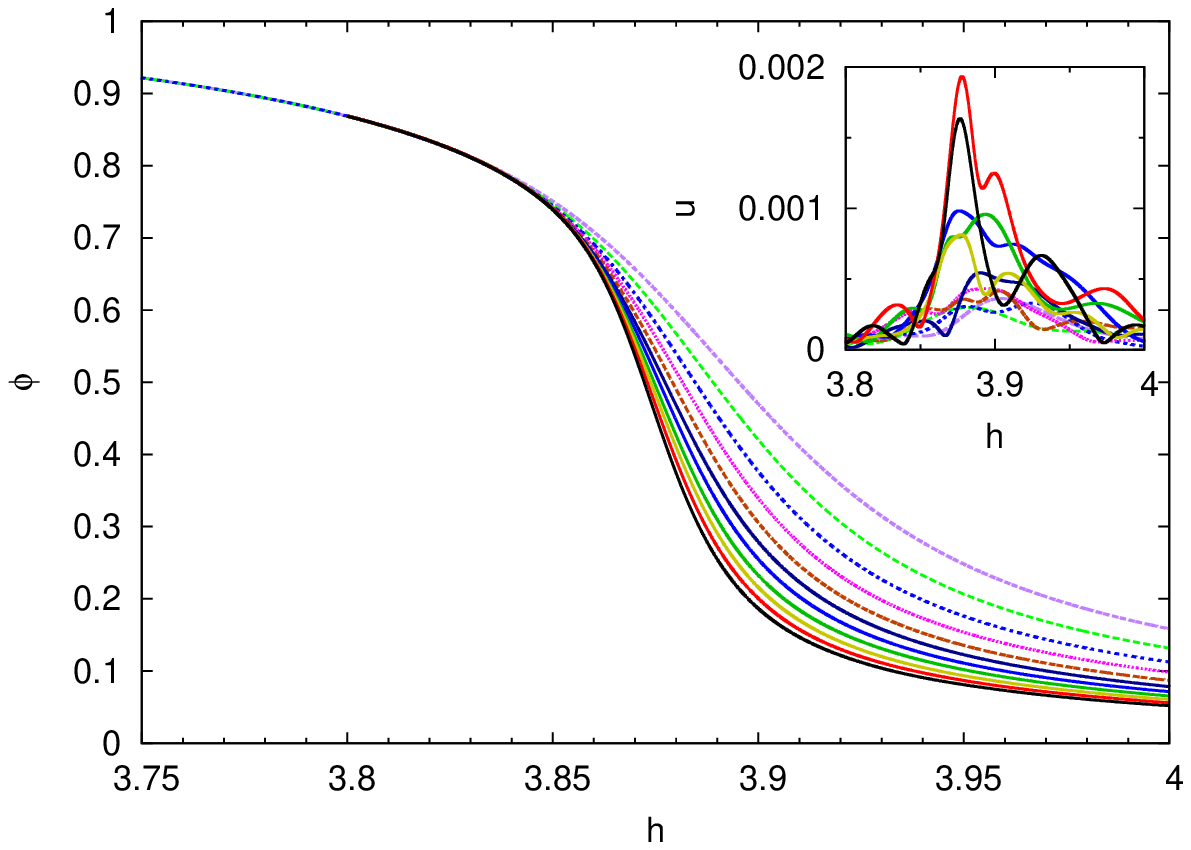}
\caption{\footnotesize{
Staggered magnetization per site as a function of $h$ at $T = 0.2$,
for $L = 10$ to 30.
The absolute uncertainties are plotted in the inset.
}
}
\label{fig:fht0_2}
\end{figure}

\begin{figure}[!h]
\centering
\includegraphics[width=0.8\textwidth]{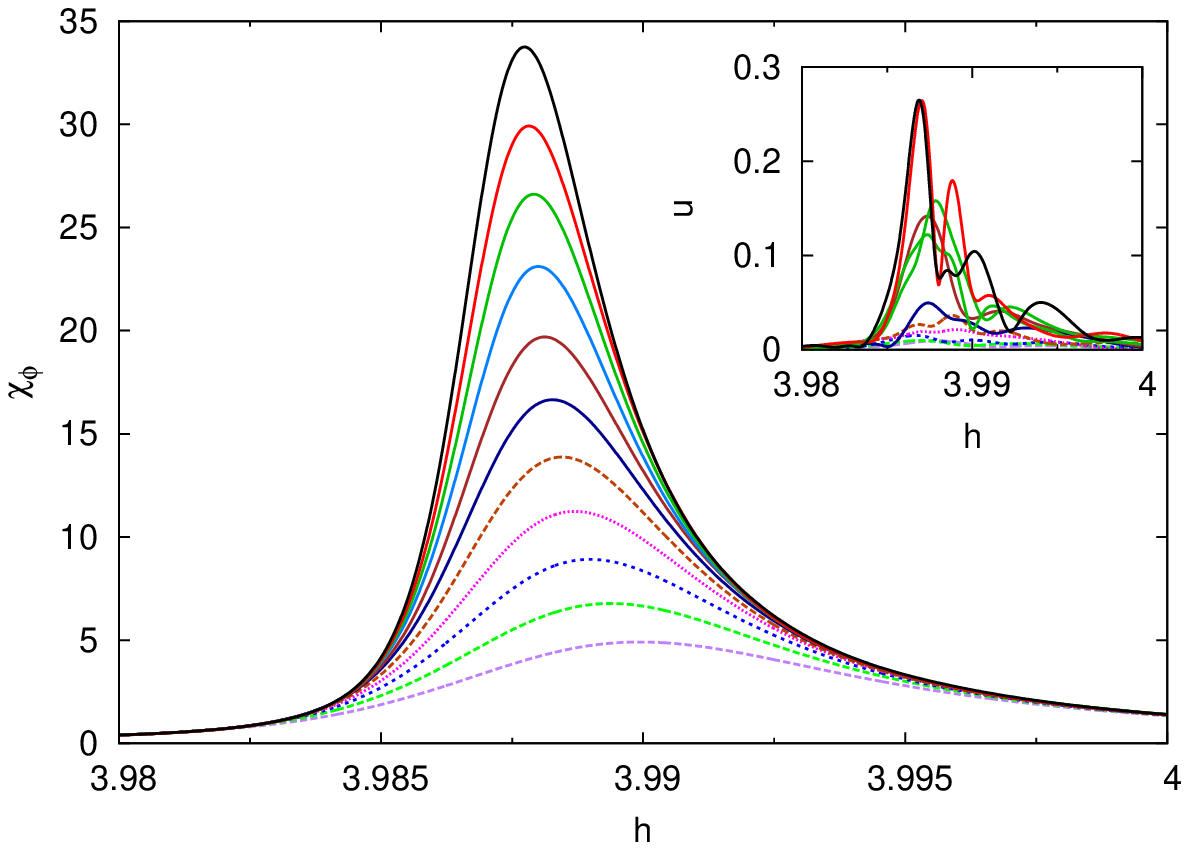}
\caption{\footnotesize{
Staggered susceptibility per site as a function of $h$ at
$T = 0.02$, for $L = 10$ to 30. The highest peaks correspond to the largest
system sizes. The absolute uncertainties are plotted in the inset.
}
}
\label{fig:xifht0_02b}
\end{figure}

In this section we present results of the AF Ising
model on the square lattice in an external field; periodic boundary
conditions are employed.
We use TES to
simulate systems of sizes $L = 10$ to 30, with $\Delta L = 2$.
To calculate the uncertainties we perform five independent studies for each system size.
We plot in Fig.~\ref{fig:fht0_2} the staggered magnetization (order parameter) per site, $\phi$,
as a function of $h$ at $T = 0.2$.
We can see that $\phi$ decreases considerably between
$h = 3.85$ and $h = 3.90$;
this behavior suggests a critical point, $h_c$,
marking a phase transition from the AF to the paramagnetic state.
Figure~\ref{fig:xifht0_02b} shows the staggered susceptibility per site,
\begin{equation}
\label{eq:chif}
\chi_\phi \equiv \frac{1}{L^2} (\langle \phi^2 \rangle - \langle \phi \rangle^2),
\end{equation}
as a function of $h$ at $T = 0.02$.
As $L$ grows the peaks tend to the critical point, $h_c$.
The specific heat per site, $c$, (not shown) has a similar behavior.

%

\subsection{Phase diagram}

\begin{figure}[ph]
\centering
\includegraphics[width=0.8\textwidth]{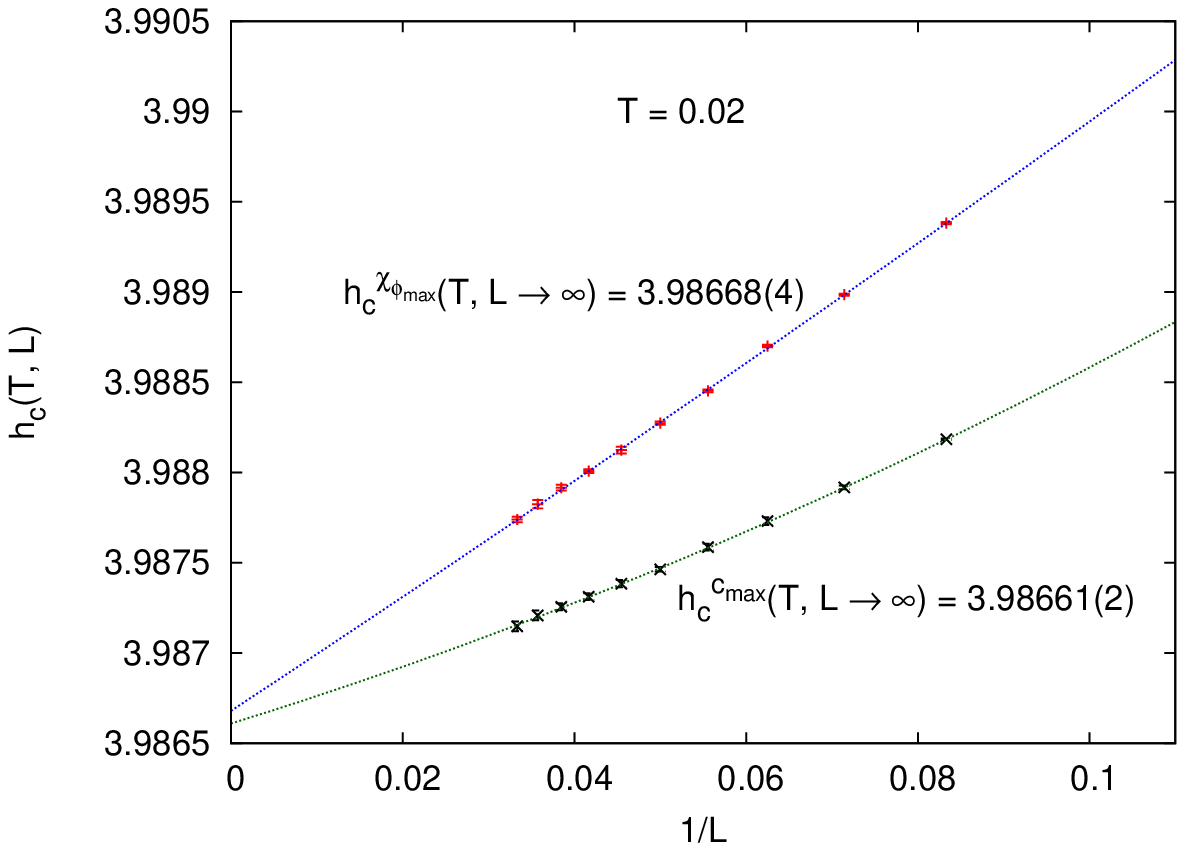}
\caption{\footnotesize{Critical field determination using
finite size scaling analysis: $\overline{h}_c(T = 0.02) = 3.98666(3)$.
}
}
\label{fig:hct0_02}
\end{figure}

\begin{figure}[ph]
\centering
\includegraphics[width=0.8\textwidth]{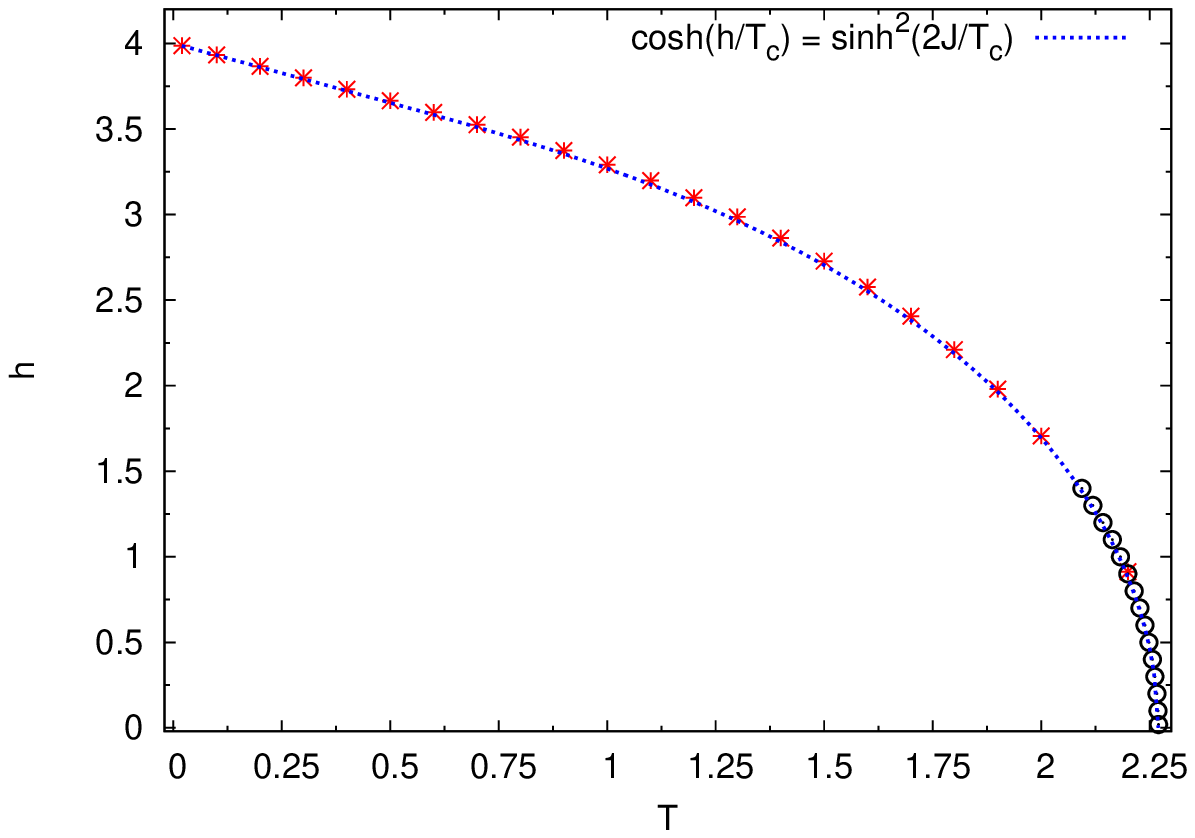}
\caption{\footnotesize{
Phase diagram of the Ising AF on the square lattice. Comparison between simulation
and the theoretical expression of M\"uller-Hartmann and Zittartz.
Asterisks denote points obtained varying $h$ with $T$ fixed; circles denote points
obtained varying $T$, with $h$ fixed. The error bars of our results are smaller than the
symbols.
}
}
\label{fig:ht_av}
\end{figure}

Using finite size scaling analysis \cite{privman90a}
we estimate the critical line, $h_c(T)$,
or, equivalently, $T_c(h)$, via the relations:

\begin{equation}
 h_c(Y_{max}, T, L) = h_c(Y_{max}, T) + a_1/L + a_2/L^2,
\end{equation}
and
\begin{equation}
 T_c(Y_{max}, h, L) = T_c(Y_{max}, h) + b_1/L + b_2/L^2,
\end{equation}
where $h_c(Y_{max},T,L)$ is the field at which $Y$ 
(the specific heat or the staggered susceptibility)
takes its maximum for a given temperature and system size;
$T_c(Y_{max},h,L)$ is defined in an analogous manner.
The estimates obtained using the maximum of $c$ and $\phi$
are averaged to yield $h_c(T)$ and $T_c(h)$;
Figure~\ref{fig:hct0_02} illustrates the procedure for estimating $h_c$,
for $T = 0.02$.
Our estimated points for the phase boundary are shown in Fig.~\ref{fig:ht_av} along
with the approximate expression derived by M\"uller-Hartmann and Zittartz \cite{hartmann77a}:
\begin{equation}
\cosh \bigg( \frac{h}{T_c} \bigg) = \sinh^2 \bigg( \frac{2 J}{T_c} \bigg).
\end{equation}
Our simulation results are in good agreement with their expression.
For the critical field, the greatest relative difference between theory
and simulation is about 0.9\%, which occurs at $T_c = 1.8$.


In Table~\ref{tab1} we compare our simulation estimates for the
critical magnetic field $h_c(T)$ with some theoretical approximations.
For temperature $T \leq 1$ we find good agreement with the estimates
of Monroe \cite{monroe01a} (whose analysis involves a free parameter $\omega$), Wu and Wu \cite{wu90a},
and Bl\"ote and Wu \cite{blote90a}, whereas there are
significant discrepancies in relation to the other approximations.
At higher temperatures, differences appear between simulation and
the predictions of Monroe, Wu and Wu, and Bl\"ote and Wu.
These may reflect a small systematic error or an underestimate of simulation
uncertainties. 
We intend to examine this issue in greater detail in future work.

\begin{table}[!h]
\centering
\caption{\bf{Comparison between our simulation estimates of $h_c(T)$ with some theoretical approximations.}}
{\begin{tabular}{@{}c|c|c|c|c|c|c|c@{}}
\hline
\hline
$T$ & TES         & Monroe   & Monroe  &  MHZ     & WW      & BW      & WK \\

  &  & $(\omega = 0.92484)$ & $(\omega = 0.93895)$ &  &  &  & \\

\hline

0.1 & 3.93304(16) &  3.93307 & 3.93318 & 3.93069 & 3.93329 & 3.93330 & 3.93372 \\

0.5 & 3.6648(8)   &  3.66506 & 3.66561 & 3.65309 & 3.66611 & 3.66614 & 3.67589 \\

1.0 & 3.2906(14)  &  3.29303 & 3.29391 & 3.26843 & 3.29200 & 3.29261 & 3.31764 \\

1.5 & 2.7258(14)  &  2.73243 & 2.73396 & 2.70401 & 2.73094 & 2.73176 & 2.75099 \\

2.0 & 1.696(2)    &  1.71629 & 1.71872 & 1.69490 & 1.71492 & 1.71499 & 1.71512 \\

\hline
\hline
\multicolumn{8}{l}{TES: Tomographic entropic sampling \cite{dickman11a}} \\
\multicolumn{8}{l}{MHZ: M\"uller-Hartmann and Zittartz \cite{hartmann77a}} \\
\multicolumn{8}{l}{WW: Wu and Wu \cite{wu90a}} \\
\multicolumn{8}{l}{BW: Bl\"ote and Wu \cite{blote90a}} \\
\multicolumn{8}{l}{WK: Wang and Kim \cite{wang97b}} \\
\end{tabular}
\label{tab1}}
\end{table}

%
\section{Conclusions}
\label{sec:conc}

The complete enumeration of CCs is of fundamental importance to study the
antiferromagnetic Ising model using tomographic entropic sampling. 
The determination of CCs for entropic
sampling of Ising models also enables us to obtain good initial estimates
for the configuration numbers $\Omega_0(n,m,L)$, using two-dimensional
linear interpolation;
the initial estimate is reasonably close to the
final estimate $\Omega_N$.
Despite the relatively small system sizes used
in this study, we obtain a good estimate for the critical line in the
temperature-magnetic field plane.
Further details on critical behavior will be published elsewhere~\cite{lourenco11a}.

%
\section*{Acknowledgments}

 We are grateful to CNPq and CAPES, Brazil, for financial support.

%

\end{document}